\documentclass[twocolumn,showpacs,preprintnumbers,amsmath,amssymb]{revtex4}
\def\brn{\begin{eqnarray*}}
\def\ern{\end{eqnarray*}}
\newcommand{\spurion}{{0 \choose 1}}
\def\rb {{\bf r}}
\def\mbs{\mbox{\boldmath$\sigma$}}

\def\sss{\scriptscriptstyle}

\def\rf#1{{(\ref{#1})}}
\def\be{\begin{equation}}
\def\ee{\end{equation}}
\def\br{\begin{eqnarray}}
\def\er{\end{eqnarray}}
\def\bc{\begin{center}}
\def\ec{\end{center}}

\def\v#1{v(#1)}

\usepackage{graphicx}% Include figure files
\usepackage{dcolumn}% Align table columns on decimal point
\usepackage{bm}% bold math

\begin{document}

\preprint{APS/123-QED}

\title{$\sigma$ meson exchange effect on nonmesonic hypernuclear weak decay observables}

\author{C. Barbero}
 \email{barbero@venus.fisica.unlp.edu.ar}
\author{A. Mariano}
\affiliation{ Departamento de F\'\i sica, Facultad de Ciencias
Exactas, Universidad Nacional de La Plata, C.C. 67, 1900 La Plata,
Argentina}

\date{\today}

\begin{abstract}
We analyze the influence of $\sigma$ meson exchange on the main
nonmesonic hypernuclear weak decay observables: the total rate,
$\Gamma_{NM}$, the neutron-to-proton branching ratio,
$\Gamma_{n/p}$, and the proton asymmetry parameter, $a_\Lambda$.
The $\sigma$ meson exchange is added to the standard
strangeness-changing weak $\Lambda N\rightarrow NN$ transition
potential, which includes the exchange of the complete
pseudoscalar and vector mesons octet ($\pi$, $\eta$, $K$, $\rho$,
$\omega$, $K^*$). Using a shell model formalism, the $\sigma$
meson weak coupling constants are adjusted to reproduce the recent
$\Gamma_{NM}$ and $\Gamma_{n/p}$ experimental data for
$^5_{\Lambda}He$. Numerical results for the remaining observables
of $^5_{\Lambda}He$ and all the observables of $^{12}_{\Lambda}C$
decays are presented. They clearly show that the addition of the
$\sigma$ meson, in spite of improving some observables values, is
not enough to reproduce simultaneously all the measurements, and
the puzzle posed by the experimental data remains unexplained.
\end{abstract}

\maketitle

The free decay of a $\Lambda$ hyperon occurs almost exclusively
through the mesonic mode, $\Lambda \rightarrow \pi N$,  emerging
the nucleon with a momentum of about $100$ MeV/c. Inside the
nuclear medium ($p_F\sim 270$ MeV/c) this mode is Pauli blocked
and, for all but the lightest $\Lambda$ hypernuclei ($A\ge 5$),
the weak decay is dominated by the nonmesonic channel, $\Lambda
N\rightarrow NN$, with enough kinetic energy to put the
two emitted nucleons above the Fermi surface. The nonmesonic
hypernuclear weak decay (NMHWD) offers us a very unique
opportunity to investigate strangeness changing weak interaction
between hadrons.
The NMHWD transitions receive contributions either from neutrons
($\Lambda n\rightarrow nn$) and protons ($\Lambda p\rightarrow
np$), with rates $\Gamma_n$ and $\Gamma_p$, respectively
($\Gamma_{NM}=\Gamma_n+\Gamma_p$, $\Gamma_{n/p}=\Gamma_n/\Gamma_p$).
Over last three decades a huge
theoretical and experimental efforts have been invested to solve
an interesting puzzle: the impossibility of the theoretical models
in reproducing at the same time experimental values of $\Gamma_{n/p}$
and $a_\Lambda$.

From the experimental side there is actually an intense activity,
as can be seen in the light of the experiments under way and/or
planned at KEK \cite{Outa00}, FINUDA \cite{Fel01} and BNL
\cite{Gill01}. The preliminary results give a $\Gamma_{n/p}$ ratio
value very close to 0.5 \cite{Gar04,Bhang05,Outa05} and the
measurements of $a_\Lambda$ favour a negative value for
$^{12}_\Lambda C$ and a positive value for $^{5}_\Lambda He$
\cite{Bhang05,Outa05,Mar05,Aji00,Aji92}. On the other hand, from
the pioneering work of Block and Dalitz \cite{Bl63} there has been
many theoretical attempts dedicated to solve the puzzle. The
earlier studies were based on the simplest model of the virtual
pion exchange \cite{Ada67}. This model naturally explains the long
range part of the two body interaction, and reproduces reasonably
well the total decay rate but badly fails in reproducing the other
observables. In order to get a better description about short
range part of the interaction it has been introduced via: (i)
models which include exchange of different combinations of other
heavier mesons, like $\eta$, $K$, $\rho$, $\omega$ and $K^*$
\cite{Mc84}--\cite{Bar05}; (ii) analysis of two nucleon stimulated
process $\Lambda NN\rightarrow NNN$ \cite{Ra95,Ra97}; (iii)
inclusion of interaction terms that violate the isospin
${\Delta}T=1/2$ rule \cite{Pa98}--\cite{Jun02}; (iv) description
of the short range baryon-baryon interaction in terms of quark
degrees of freedom \cite{In98,Sa00}; (v) correlated (in the form
of $\sigma$ and $\rho$ mesons) and uncorrelated two-pion exchanges
\cite{Shm94}--\cite{Sas05}. We emphasize that none of these models
give a fully satisfactory description of all the NMHWD observables
simultaneously, in spite that consistent (though not sufficient)
increase of $\Gamma_{n/p}$ have been found. Only Jun \cite{Jun02}
was able so far to reproduce well the $\Gamma_{NM}$ and
$\Gamma_{n/p}$ (but not $a_\Lambda$) data employing, in addition
to the one-pion exchange, an entirely phenomenological four-baryon
point interaction for the short range part, including the $\Delta
T=3/2$ contribution as well, and fixing the model coupling
constants. Concerning the proton asymmetry parameter $a_\Lambda$,
all existing calculations based on strict one-meson exchange
models \cite{Sa00,Du96,Pa97,Pa01,Bar03,Ito02,Sas02,Bar05} find
values between $-0.73$ and $-0.19$ for $^5_\Lambda He$
\cite{Alb02,Alb04} and, when results are available in the same
model, very similar values for $^{12}_\Lambda C$. A recent attempt
\cite{Alb04} to explain these discrepancies with the experimental
data by means of the FSI effect has failed because, in spite of
attenuating the $a_\Lambda$ value, its effect does not reverse the
sign of the parameter. Only two recent theoretical calculations
show some agreement with the experimental data: (i) a first
application of effective field theory (EFT) to nonmesonic decay
\cite{Par04}, and (ii) a very recent extension of the direct-quark
interaction model to include $\sigma$ meson exchange \cite{Sas05}.
However, in both cases, a nuclear matter formalism (which does not
includes the full contribution of transitions coming from nucleon
states beyond the $s$-shell) is used, and could give therefore a
limited description for $p$-shell hypernuclei or, even worse, for
heavier ones.

The EFT approach from Ref. \cite{Par04} suggests, in order to
reproduce the NMHWD data, that the microscopic models should be
supplemented by isospin and spin independent central interactions.
Motivated by this fact and to set a more detailed description than
previous calculations in nuclear matter \cite{Sas05}, we analyze
the influence of the scalar-isoscalar $\sigma$ meson exchange over
the NMHWD. We will work within the shell model (SM) formalism
developed in Refs. \cite{Ba02}--\cite{Bar05}, which explicitly
includes the contribution of transitions originated from states
beyond the $s$-shell. Therefore, within our model the $\sigma$
meson is added to the standard strangeness-changing weak $\Lambda
N\rightarrow NN$ transition potential already including the
exchange of the complete pseudoscalar and vector mesons octet
($\pi$, $\eta$, $K$, $\rho$, $\omega$, $K^*$). This model will be
referenced as $OCT+S$, to differentiate with our previous
$\pi+\eta+K+\rho+\omega+K^*$ model, designed as $OCT$.

For the pseudoscalar and vector mesons octet, the weak (W) and a
strong (S) vertices in the $\Lambda N\rightarrow NN$ decay will be
described by means of the same interaction Hamiltonians given in
Ref. \cite{Pa97}, as it has just been adopted in Refs.
\cite{Ba02}--\cite{Bar05}. This lead to the $OCT$ exchange
potential \be V^{OCT}(\rb) = \sum_{i=\pi,\eta,K,\rho,\omega,K^*}
\bar{V}_i^{(0)}(\rb ), \label{1}\ee with $\bar{V}_i^{(0)}(\rb )$
defined in Ref. \cite{Bar03}. In this equation we have neglected
all kinematical and first-order nonlocality corrections because,
as it has been extensively discussed in Ref. \cite{Bar03},
although their effects can be very important for particular
transitions, they do not affect too much the main decay
observables. For the $\sigma$ meson we assume weak (W) and strong
(S) coupling Hamiltonians of the form (\cite{Sas05}, Eq. (17))
\br
{\cal H}^{\sss W}_{{\sss \Lambda N}\sigma} &=&G_{\sss F}\mu_\pi^2
\bar{\psi}_{\sss N}(A_\sigma+B_\sigma\gamma_5)
\phi_\sigma\spurion\psi_{\sss \Lambda},\nonumber\\
{\cal H}^{\sss S}_{{\sss NN}\sigma} &=&g_{{\sss NN}\sigma}
\bar{\psi}_{\sss N}\phi_\sigma\psi_{\sss N},
\label{2}\er
where ${\psi}_{\sss N}$ and $\psi_{\sss \Lambda}$ are the baryon
fields, $\phi_\sigma$ is the meson field, and $A_\sigma$ and
$B_\sigma$ are the weak parity conserving (PC) and parity
violating (PV) coupling constants, respectively, which will be
considered as adjustable parameters of the model. For the strong
coupling constant we will assume the value $g_{{\sss
NN}\sigma}\simeq g_{{\sss NN}\pi}=13.3$. The resulting
nonrelativistic one-sigma-exchange potential is
\be
V_\sigma(\rb)=G_{\sss F}\mu_\pi^2 g_{{\sss NN}\sigma} \left[
A_\sigma f_C(r,\mu_\sigma)
-i \frac{B_\sigma}{2\bar{M}} f_V(r,\mu_\sigma)
\mbs_1\cdot\hat{\rb}\right],
\label{3}\ee
where $\mu_\sigma$ is the $\sigma$ meson mass, ${\bar
{M}}=(M_\Lambda+M)/2$ with $M$ and $M_\Lambda$ being the nucleon
and $\Lambda$ masses, respectively, and all the remaining notation
has the same meaning as in Ref. \cite{Bar03}. Thus, we have
\be
V^{OCT+S}(\rb) = V^{OCT}(\rb)+V_\sigma(\rb).
\label{4}\ee
\begin{table}[h]
\begin{center}
\caption {Parity conserving and parity violating constants for
neutron and proton NMHWD rates of $^5_\Lambda He$.}
\label{tab1}
\begin{tabular}{c|ccc}
&&$\mu_\sigma$ [MeV]&\\
&$550$&$650$&$750$\\
\hline
&&&\\
$a_n^{PC}$&$0.0096 $&$0.0032 $&$0.0010 $\\
$b_n^{PC}$&$1.0806 $&$1.8692 $&$3.4141 $\\
$c_n^{PC}$&$0.0000 $&$0.0000 $&$0.0000 $\\
$a_n^{PV}$&$0.0022 $&$0.0008 $&$0.0003 $\\
$b_n^{PV}$&$7.3380 $&$12.0755 $&$21.0396 $\\
$c_n^{PV}$&$0.0508 $&$0.0508 $&$0.0509 $\\
$a_p^{PC}$&$0.0192 $&$0.0064 $&$0.0019 $\\
$b_p^{PC}$&$1.3623 $&$2.3588 $&$4.2979 $\\
$c_p^{PC}$&$0.1422 $&$0.1422 $&$0.1421 $\\
$a_p^{PV}$&$0.0014 $&$0.0005 $&$0.0002 $\\
$b_p^{PV}$&$12.4826 $&$20.5396 $&$35.8000 $\\
$c_p^{PV}$&$0.1398 $&$0.1398 $&$0.1399 $\\
\end{tabular} \end{center}
\end{table}
\vspace{-0.5cm}

Using the SM formalism developed in Refs. \cite{Ba02,Bar03} we
evaluate the partial neutron $\Gamma_n$ and proton $\Gamma_p$
induced decay rates for $^5_\Lambda He$. We include finite nucleon
size effect and short range correlations following Ref.
\cite{Ba02}, and use a cutoff value $\Lambda_\sigma=1200$ MeV for
the $\sigma$ meson. The partial decay rates read
\be
\Gamma_i=a_i^{PC}(A_\sigma-b_i^{PC})^2+c_i^{PC}+a_i^{PV}(B_\sigma-b_i^{PV})^2
+c_i^{PV},
\label{5}\ee
where $i=n,p$ and the coefficients $a_i^{X}$, $b_i^{X}$ and
$c_i^{X}$ (which come from SM matrix elements involving the
$\sigma$ exchange) are listed in Table \ref{tab1} for different
values of the $\sigma$ meson mass. Solving now the equations
system obtained fixing $\Gamma_n+\Gamma_p$ and $\Gamma_n/\Gamma_p$
to the $^5_\Lambda He$ experimental central values from Ref.
\cite{Outa05} (shown in Table \ref{tab2}) we obtain the following
two sets of solutions for $A_\sigma$ and $B_\sigma$:
\[
A_\sigma=0.67, B_\sigma=13.39 \mbox{ and }A_\sigma=2.08, B_\sigma=13.08,
\]
for $\mu_\sigma=550$ MeV,
\[
A_\sigma=1.16, B_\sigma=22.03 \mbox{ and }A_\sigma=3.60, B_\sigma=21.52,
\]
for $\mu_\sigma=650$ MeV and
\[
A_\sigma=2.12, B_\sigma=38.39 \mbox{ and }A_\sigma=6.56, B_\sigma=37.50,
\]
for $\mu_\sigma=750$ MeV, respectively.
\begin{table}[h]
\begin{center}
\caption {Experimental data for the NMHWD of $^5_\Lambda He$ and
$^{12}_\Lambda C$ (total width is in units of
$\Gamma_0 = 2.50 \times 10^{-6}$ eV).}
\label{tab2}
\begin{tabular}{ccc}
\hline \hline
Observable&$^5_\Lambda He$&$^{12}_\Lambda C$\\
\hline
$\Gamma_{NM}$&$0.424\pm 0.024$ \cite{Outa05}&$0.940\pm 0.035$ \cite{Outa05}\\
&$0.41\pm 0.14  $ \cite{Szy91}&$1.14\pm 0.2$ \cite{Szy91}\\
\hline
$\Gamma_{n/p}$&$0.39\pm 0.11$ \cite{Gar04}&$-$\\
&$(0.45-0.51)\pm 0.15$ \cite{Bhang05}&$0.87\pm0.09\pm 0.21$ \cite{Bhang05}\\
&$0.45\pm 0.11\pm 0.03$ \cite{Outa05}&$0.56\pm 0.12\pm 0.04$ \cite{Outa05}\\
&$0.93\pm 0.55$ \cite{Szy91}&$1.33^{+0.12}_{-0.81}$ \cite{Szy91}\\
\hline
$a_\Lambda$&$0.07\pm 0.08^{+0.08}_{-0.00}$ \cite{Bhang05}&$-0.24\pm 0.26^{+0.08}_{-0.00}$ \cite{Bhang05}\\
&$0.11\pm 0.08\pm 0.04$ \cite{Outa05,Mar05}&$-0.20\pm 0.26\pm 0.04$ \cite{Outa05,Mar05}\\
&$0.24\pm 0.22$ \cite{Aji00}&$-$\\
\hline \hline
\end{tabular} \end{center}
\end{table}

Following, we evaluate $a_\Lambda$ for $^5_\Lambda He$ using the
SM formalism recently developed in Ref. \cite{Bar05}. We get
\be
a_\Lambda=(k_1A_\sigma B_\sigma+k_2A_\sigma+k_3B_\sigma+k_4)/\Gamma_p,
\label{6}\ee
where the constants $k_i$ (containing the SM information) will be
given below for a particular value of the $\sigma$ meson mass.
Numerical results obtained with the fixed values of the weak
coupling constants have been collected in Table \ref{tab3},
together with the predicted values of all the observables for the
p-shell $^{12}_\Lambda C$ hypernucleus.

\begin{table}[h]
\begin{center}
\caption {Hypernuclear weak decay observables for the two sets of
coupling constants, for each $\mu_\sigma$ ($\Gamma_{NM}$ is given
in units of $\Gamma_0 = 2.50 \times 10^{-6}$ eV).} \label{tab3}
\begin{tabular}{ccc|ccc|ccc}
$\mu_\sigma$ [MeV]&$A_\sigma$&$B_\sigma$&&$^{5}_\Lambda He$&&&$^{12}_\Lambda C$&\\
\hline
&&&$\Gamma_{NM}$&$\Gamma_{n/p}$&$a_\Lambda$&$\Gamma_{NM}$&$\Gamma_{n/p}$&$a_\Lambda$\\
\hline
$-$&$0$&$0$&$0.720$&$0.329$&$-0.507$&$1.166$&$0.267$&$-0.508$\\
$550$&$0.67$&$13.39$&$0.424$&$0.450$&$-0.326$&$0.776 $&$0.427 $&$-0.322 $\\
$550$&$2.08$&$13.08$&$0.424$&$0.450$&$-0.362$&$0.793 $&$0.435 $&$-0.348 $\\
$650$&$1.16$&$22.03$&$0.424$&$0.450$&$-0.326$&$0.771$&$0.421$&$-0.323$\\
$650$&$3.60$&$21.52$&$0.424$&$0.450$&$-0.362$&$0.786$&$0.428$&$-0.344$\\
$750$&$2.12$&$38.39$&$0.424$&$0.450$&$-0.326$&$0.767$&$0.416$&$-0.324$\\
$750$&$6.56$&$37.50$&$0.424$&$0.450$&$-0.362$&$0.780$&$0.423$&$-0.342$\\
\end{tabular} \end{center}
\end{table}

They clearly show that our results are not sensible to the
particular $\mu_\sigma$, $A_\sigma$ and  $B_\sigma$ selected set,
because in all cases our $OCT+S$ model predicts the values
$a_\Lambda(^5_{\Lambda}He)\simeq -0.34$,
$\Gamma_{NM}(^{12}_{\Lambda}C)\simeq 0.78$,
$\Gamma_{n/p}(^{12}_{\Lambda}C)\simeq 0.43$ and
$a_\Lambda(^{12}_{\Lambda}C)\simeq -0.33$. This evidences the fact
that these three variables are strongly correlated between them
and shows the stability of our results, as was the case with the
inclusion of the $\sigma$ meson in $\pi N$ scattering calculations
\cite{Ale01}. A straightforward comparison of $OCT+S$ results with
the $OCT$ ones shows that the inclusion of the $\sigma$ meson
reduces the results for $\Gamma_{NM}$ by $\sim 40$ \%. In
addition, the neutron to proton branching ratio $\Gamma_{n/p}$ is
increased $\sim 36$ \%. These last two effects are obtained since
the $\sigma$ meson strongly reduces the PV contribution of the
proton induced decay, as can be seen from Eq. \rf{5} and from the
fact that $B_\sigma\simeq b_p^{PV}$. This effect can also be
observed from Table \ref{tab4}.

\begin{table}[h]
\begin{center}
\caption {Comparison of partial rates contributions within OCT
and OCT+S models for $\mu_\sigma=550$ MeV, $A_\sigma=0.67$ and
$B_\sigma=13.39$ (in units of $\Gamma_0 = 2.50 \times 10^{-6}$ eV).}
\label{tab4}
\bigskip
\begin{tabular}{cc|cccc}
\hline
&Model&$\Gamma_{n}^{PC}$&$\Gamma_{n}^{PV}$&$\Gamma_{p}^{PC}$&$\Gamma_{p}^{PV}$\\
\hline
$^{5}_\Lambda He$&OCT&$0.011$&$0.167$&$0.178$&$0.364$\\
&OCT+S&$0.002$&$0.130$&$0.151$&$0.141$\\
\hline
$^{12}_\Lambda C$&OCT&$0.024$&$0.222$&$0.296$&$0.624$\\
&OCT+S&$0.005$&$0.227$&$0.251$&$0.293$\\
\hline
\end{tabular} \end{center}
\end{table}

However, in spite this meson helps to bring near zero the
asymmetry $a_\Lambda$ (increasing it $\sim 35$ \%), its effect is
not enough to reverse the sign of the parameter, as required in
the case of $^5_\Lambda He$ hypernucleus. This can be understood
noting that, for example for $\mu_\sigma=550$ MeV, we have
$k_1=-0.010$, $k_2=0.127$, $k_3=0.014$ and $k_4=-0.275$. Thus, the
strong $B_\sigma$ and weak $A_\sigma$ coupling constants required
to increase the $\Gamma_{n/p}$ ratio, makes not possible to
reverse the $a_\Lambda$ sign, at least within the implemented
hypernuclear model. For the p-shell $^{12}_\Lambda C$ hypernucleus
our result is consistent with the experimental data, which have
large error bars.

\begin{figure}
\begin{center}
\vspace{-0.5cm}
\includegraphics[height=15.3cm,width=9.cm]{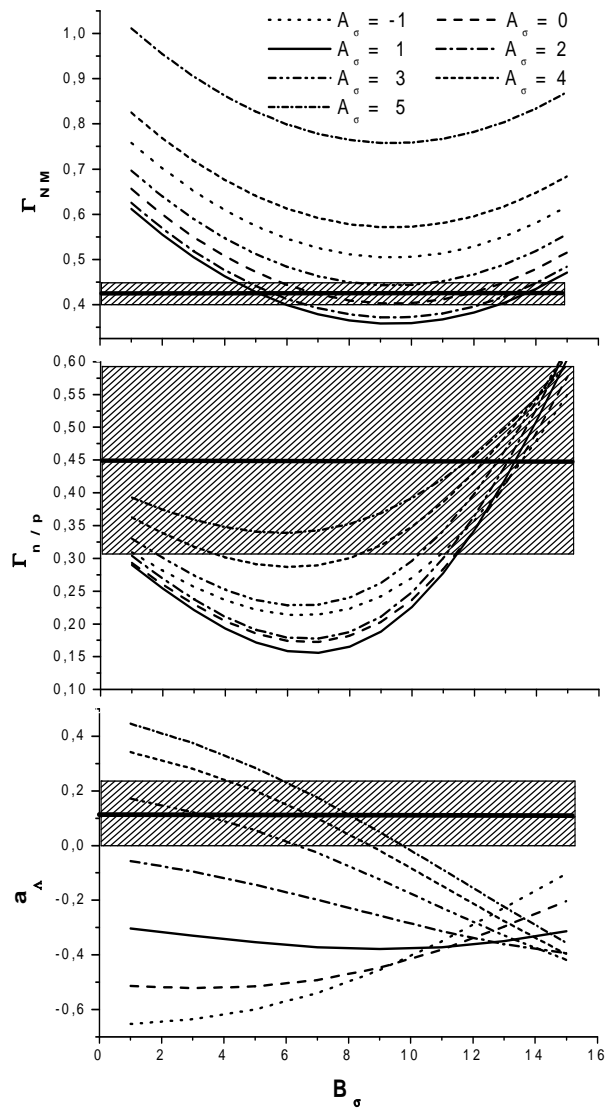}
\vspace{-0.5cm}
\caption{\label{fig1} $B_\sigma$ dependence of the $^5_\Lambda He$
decay observables, for some particular fixed values of $A_\sigma$
and for $\mu_\sigma=550$ MeV. The shaded region stands for the
experimental values \cite{Outa05} with error bars.}
\end{center}
\end{figure}

In order to illustrate the $A_\sigma$ and $B_\sigma$ dependence of
the NMHWD observables, we present in Fig. \ref{fig1} our results
for $^5_\Lambda He$, using $\mu_\sigma=550$ MeV. This figure shows
that within our $OCT+S$ model a bigger value for $A_\sigma$ and a
smaller for $B_\sigma$ are necessary for reversing the asymmetry
sign; however, this leads to overestimate the total rate
$\Gamma_{NM}$ and strongly underestimate the ratio $\Gamma_{n/p}$.
Thus, the $OCT+S$ model implemented with the SM formalism shows
that the inclusion of the $\sigma$ meson is not enough to
reproduce simultaneously all the data, specially the positive
proton asymmetry value for $^{5}_\Lambda He$.

Summarizing, we have made an analysis of the $\sigma$ meson
contribution to the NMHWD observables in the framework of a SM
formalism \cite{Ba02}--\cite{Bar05}, which includes carefully the
effect of the p-shell state decays, both in the partial decay
rates as in the asymmetry parameter evaluation. From our results
we conclude that, in spite that the $\sigma$ meson contributes
reducing the total decay width $\Gamma_{NM}$ and increasing the
neutron to proton branching  ratio $\Gamma_{n/p}$, it is not
enough to reproduce the more recent experimental data on the
asymmetry. In fact, in the case of the s-shell $^5_\Lambda He$
hypernucleus, is not possible to reverse the sign of $a_\Lambda$,
while for the p-shell $^{12}_\Lambda C$ hypernucleus we have
obtained a better agreement. Another remarkable characteristic
shown by our results is to corroborate the theoretical expectation
that $a_\Lambda$ should have only a moderate dependence on the
particular hypernucleus considered, as can be seen from Table
\ref{tab3}. We suggest two possible alternatives for future work:
(i) additional degrees of freedom should be added in our SM
formalism to improve the description of the experimental data, for
example $\Delta I=3/2$ contributions in the same footing of Ref.
\cite{Sas05} (but now in a finite nucleus scheme) where a better
agreement has been obtained for $^{5}_\Lambda He$; (ii) to
introduce a different decay mechanism for s and p-shell
hypernuclei, beyond the Born approximation. Finally, in spite of
the uncertainties on the $\sigma$ meson properties, from the
results presented in Table \ref{tab3} we have observed that our
results are almost independent of the particular choice for the
$\sigma$ mass, and we have clearly shown that the effects of an
isoscalar central spin-independent interaction included through
the $\sigma$ meson would not be enough to reproduce simultaneously
all the NMHWD observables, without additional changes in the
nuclear model and/or the decay mechanism.

The authors are fellows of the CONICET (Argentina). C.B.
acknowledge the support of CONICET under Grant No. 666.

\end{document}